\documentclass[aps,prb,reprint,groupedaddress]{revtex4-1}
\usepackage{amsmath}
\usepackage{amssymb}
\usepackage{graphicx}
\usepackage{bm}

\begin{document}

\title{First Principles Quantum Transport with Electron-vibration Interactions:\\
A Maximally Localized Wannier Function Approach}

\author{Sejoong Kim}
\altaffiliation[Present address: ]{School of Computational Sciences, Korea Institute for Advanced Study, Seoul 130-722, Korea}
\affiliation{Department of Physics, Massachusetts Institute of Technology, Cambridge, Massachusetts 02139, USA}

\author{Nicola Marzari}
\affiliation{Theory and Simulation of Materials (THEOS), \'{E}cole Polytechnique F\'{e}d\'{e}rale de Lausanne, 1015 Lausanne, Switzerland}

\begin{abstract}
We present an \emph{ab initio} inelastic quantum transport approach based on maximally localized Wannier
functions. Electronic-structure properties are calculated with density-functional theory in a planewave basis, 
and electron-vibration coupling strengths and vibrational properties are determined with density-functional perturbation theory. 
Vibration-induced inelastic transport properties are calculated with non-equilibrium Green's function techniques, which are
based on localized orbitals. For this purpose we construct maximally localized Wannier functions. 
Our formalism is applied to investigate inelastic transport in a benzene molecular junction connected to 
mono-atomic carbon chains. In this benchmark system the electron-vibration self-energy is calculated either in the self-consistent Born 
approximation or by lowest-order perturbation theory. It is observed that upward and downward conductance steps occur,
which can be understood using multi-eigenchannel scattering theory and symmetry conditions. In a second example where the mono-atomic
carbon chain electrode is replaced by a $(3,3)$ carbon nanotube, we focus on the non-equilibrium vibration populations driven by the
conducting electrons using a semi-classical rate equation. 

\end{abstract}

\maketitle

\section{introduction}
Molecular electronic devices have been intensively studied for the past decade, 
having been regarded as candidates to replace silicon-based electronic devices. 
Since Aviram and Ratner proposed the concept of the first molecular rectifier in 1974\cite{1974Aviram}, a variety of 
molecular devices have been suggested. 
In addition, thanks to advancement in experimental fabrication and measurement techniques, 
electronic currents through molecules have been experimentally measured and investigated
\cite{1997Reed, 1998Scheer, 1998Yanson, 2002Smit, 2006Venkataraman, 2003Xu}. 
Despite these efforts, there are still many issues in the practical realization of molecular electronic devices. 

Understanding interactions between conducting electrons and molecular vibrations is 
one of the key issues to address for future applications. 
Vibrational excitations due to the scattering of conducting electrons can change 
molecular configurations and attachment geometries, 
affecting the functionality and performance of the molecule-based devices or in turn
backscatter electrons, impeding flow.The worst scenario is that local heating effects might ultimately break down the junction. 

Vibration-induced inelastic transport has been theoretically investigated following two directions. 
The first one is to study simple model Hamiltonians, e.g. a single electronic level coupled to a single phonon
mode, which is known as the Anderson-Holstein model\cite{2005Koch, 2006Koch1, 2006Koch2, 2004Cornaglia, 2006Galperin}. 
Based on this simplified model, many novel and interesting transport properties have been
predicted and investigated. However, the model used in this approach
is too simplified to provide detailed and accurate theoretical data that quantitatively explain experimental results.

On the other hand, a quantitative and computational approach based on density
functional theory (DFT) offers the chance to describe a realistic system accurately without
any adjustable parameters\cite{1964PRHohenberg, 1965PRKohn}. 
In particular, using DFT one can calculate equilibrium geometries, electronic couplings, normal modes, and electron-vibration interactions. 

Transport theories can then be combined with DFT, and several approaches have been proposed\cite{2003Chen, 2004Chen, 2005Chen, 2005Jiang}. 
In particular, a non-equilibrium Green’s Function (NEGF) formulation, in combination with DFT, and commonly called DFT-NEGF, has been
widely used in ab initio quantum transport problems\cite{2001PRBTaylor, 2002PRBBrandbyge, 2002PRBPalacios}. 
This approach is more powerful than other methods in that it can tackle not just the emergence of electron-vibration
interactions, but any other type of interactions. DFT-NEGF has been successfully applied to elastic quantum transport for both zero-bias\cite{2005YSLee, 2006YSLee} 
and finite-bias cases\cite{2001PRBTaylor, 2002PRBBrandbyge, 2002PRBPalacios}, and 
recently it has been extended to include interaction effects like electron-vibration
interactions\cite{2004Frederiksen, 2007Frederiksen}.

DFT-NEGF requires to use an atomic-like localized basis, because the device should be spatially divided into two electrodes and a molecular conductor. 
For this reason most of DFT-NEGF calculation packages have been implemented using localized basis set. 
However, from a computational viewpoint, it is known that a planewave-based DFT calculation can provide an accurate description of electronic states systematically, 
and in particular can describe electronic states which have considerable spread in vacuum, where localized orbitals are absent. 
Furthermore, while basis functions used in localized-basis calculations are determined depending on types
of atoms and the chemistry of the system, a planewave basis can describe any given system without making any further assumption. 

However, periodic-boundary conditions planewave calculations are not suitable to DFT-NEGF calculations. 
For this, maximally localized Wannier functions (MLWF), as proposed by Marzari and
Vanderbilt\cite{1997Marzari}, provide the formal and algorithmic formulation for a transformaton between delocalized and localized orbitals. 
Since the Wannier transformation is an exact unitary mapping, one can construct a minimal set of atomic-like localized functions within an energy window of interest without losing
the accuray of planewave-based DFT calculations, and a MLWF approach to quantum transport has been very successfully applied to zero-bias quantum conductance calculations\cite{2005YSLee, 2006YSLee}. 
The next step to develop a MLWF approach to quantum transport\cite{2005YSLee, 2006YSLee, 2011Shelley, 2011Elise} is to
include interaction effects on transport properties. In this paper we focus on extending
MLWF-based ab initio quantum transport calculations to investigate electron-vibration interaction effects on molecular junctions.

This paper is organized as follows. In Sec.\ref{section2} we review first-principles electronic
structure calculations, especially focusing on (1) density-functional perturbation theory
to calculate vibrational properties and electron-vibrational interaction, and (2) 
the transformation of electron-vibration interactions from a plane-wave basis to a maximally localized
Wannier function basis. In Sec.\ref{section3} a quantum transport theory based on non-equilibrium
Green's functions and diagrammatic perturbation theory is presented. First we discuss electrode-conductor-electrode
system partitioning and the calculation of lead self-energies. Then we review diagrammatic expansion schemes for electron-vibration
interactions in the self-consistent Born approximation and lowest-order perturbation theory. Inelastic transport properties
such as finite-bias electronic current and power transfer are calculated within the Meir-Wingreen transport formalism\cite{MeirWingreen1}. 
Finally we present a semi-classical rate equation to determine nonequilibrium vibrational populations in the presence of interactions with conducting electrons and coupling to bulk vibrations in the electrodes. 
In Sec.\ref{section4} application results and further analysis are presented.

\section{Electronic Structure Methods}
\label{section2}
\subsection{Vibrational Properties: Density-Functional Perturbation Theory}
The electronic structure of a given system is calculated within the DFT framework. 
The ground state charge density and Bloch wavefunctions are determined by solving the Kohn-Sham equations\cite{2004Martin},
\begin{equation}
\left[-\frac{\hbar^2}{2m} \nabla^2 + V_{KS} \right] |\psi_{nk}\rangle = \varepsilon_{nk} |\psi_{nk}\rangle,
\end{equation}
where 
\begin{equation}
V_{KS} = e^2\int \frac{n(\mathbf{r}^\prime)}{|\mathbf{r}-\mathbf{r}^\prime|}d\mathbf{r}^\prime + v_{xc}(\mathbf{r}) + V_{ion}(\mathbf{r}).
\end{equation}
Here $v_{xc}=\frac{\delta E_{xc}}{\delta n}$ is the exchange-correlation potential, and $V_{ion}(\mathbf{r})$ is the ionic core potential described by pseudopotentials. 
We solve the Kohn-Sham equations in a planewave basis, as implemented in the \textsc{quantum-espresso} distribution\cite{2009JPCMGiannozzi}. 

Vibrational properties and electron-vibration interactions are determined within density-functional perturbation Theory (DFPT)\cite{2001Baroni}. 
Vibrational spectra and the corresponding normal modes are obtained from the interatomic force constants, which are
the second derivatives of the Born-Oppenheimer total energy surface $E(\{\mathbf{R}\})$ with respect to displacements of the ions, 
\begin{eqnarray}
C_{\alpha i,\beta j} &=& \left.\frac{\partial^2 E}{\partial R_{\alpha i} \partial R_{\beta j}}\right|_{0} = C^{ion}_{\alpha i, \beta j} + C^{elec}_{\alpha i, \beta j},
\end{eqnarray}
where $i(j)$ represents the $i(j)$th atom of the unit cell, and $\alpha(\beta)$ indicates the Cartesian component of the displacements. 
While $C^{ion}_{\alpha i, \beta j}$ is the contribution of the ion-ion interaction potential, $C^{elec}_{\alpha i, \beta j}$ is the second derivative of electron-electron
and electron-ion interactions. From the Hellmann-Feynman theorem, one can obtain
\begin{eqnarray}
C^{elec}_{\alpha i, \beta j} = \int \left[\frac{\partial n(\mathbf{r})}{\partial R_{\alpha i}} \frac{\partial V_{ion}(\mathbf{r})}{\partial R_{\beta j}}+n(\mathbf{r}) \frac{\partial^2 V_{ion}(\mathbf{r})}{\partial R_{\alpha i} \partial R_{\beta j}} \right] d\mathbf{r}
\end{eqnarray}

The key quantity to obtain interatomic force constants is the linear variation of the charge density $n(\mathbf{r})$ with respect to ionic displacements. In the Kohn-Sham mapping, 
the linear response of the charge density $\Delta n(\mathbf{r})$ can be calculated from the Kohn-Sham orbital variation $|\Delta \psi_{i}\rangle$. 
From first-order perturbation theory, one can derive the equation for $|\Delta \psi_{i}\rangle$ \cite{1987Baroni}:
\begin{equation}
\left[-\frac{\hbar^2}{2m} \nabla^2 + V_{KS} -\varepsilon_{i} \right] |\Delta \psi_{i}\rangle = -\left[\Delta V_{KS} - \Delta \varepsilon_{i}\right]|\psi_{i}\rangle,
\end{equation}
where
\begin{equation}
\Delta V_{KS}(\mathbf{r}) = \Delta V_{ion}(\mathbf{r}) + e^2 \int \frac{\Delta n(\mathbf{r}^\prime)}{|\mathbf{r}-\mathbf{r}^\prime|}d\mathbf{r}^\prime + \left.\frac{dv_{xc}}{dn}\right|_0 \Delta n(\mathbf{r}).
\end{equation}
Here $\Delta \varepsilon_{i} = \langle \psi_{i} | \Delta V_{KS} | \psi_{i} \rangle$ is the first-order correction to the Kohn-Sham energy eigenvalue $\varepsilon_{i}$. 
Note that these equations should be self-consistently solved since the equations that determine $|\Delta \psi_{i}\rangle$ depends on the linear response in the charge density $\Delta n(\mathbf{r})$.
The vibrational frequencies $\omega$ of the system can be calculated by solving the following eigenvalue equation:
\begin{equation}
\textrm{det} \left| \frac{C_{\alpha i,\beta j}}{\sqrt{M_i M_j}} - \omega^2 \right|=0.
\end{equation}

The electron-vibration interaction can be written in a second quantized form as follows\cite{2005Wierzbowska}:
\begin{equation}
\mathcal{H}_{el-vib} = \sum_{\mathbf{k}\mathbf{q}\lambda}\sum_{mn} g_{\mathbf{k}+\mathbf{q},\mathbf{k}}^{\mathbf{q}\lambda, mn} a^{\dagger m}_{\mathbf{k}+\mathbf{q}} a_{\mathbf{k}}^{n} \left(b^{\dagger}_{-\mathbf{q}\lambda} + b_{\mathbf{q}\lambda} \right),
\end{equation}
where $a_{\mathbf{k}}^{\dagger n}$ ($a_{\mathbf{k}}^{n}$) is the electron creation (annhilation) operator for Bloch state $|\psi_{n\mathbf{k}}\rangle$. 
Similarly $b_{\mathbf{q}\lambda}^{\dagger}$ ($b_{\mathbf{q}\lambda}$) is the creation (annhilation) operator for the phonon of the vibrational mode $\lambda$ with 
the energy $\omega_{\mathbf{q}\lambda}$ at wave vector $\mathbf{q}$, and $g_{\mathbf{k}+\mathbf{q},\mathbf{k}}^{\mathbf{q}\lambda, mn}$ is the electron-phonon coupling 
matrix element. Once the linear response of the charge density is computed from DFPT, the electron-phonon coupling matrix can be calculated from the
derivative of the self-consistent Kohn-Sham potential $\Delta v_{KS}$ as follows:
\begin{equation}
g_{\mathbf{k}+\mathbf{q},\mathbf{k}}^{\mathbf{q}\lambda, mn} = \left(\frac{\hbar}{2\omega_{\mathbf{q}\lambda}} \right)^{1/2} 
\langle \psi_{\mathbf{k}+\mathbf{q},m}| \Delta V_{KS}^{\mathbf{q}\lambda} | \psi_{\mathbf{k},n}\rangle,
\end{equation}
where $\psi_{\mathbf{k},n}$ is the $n$th Kohn-Sham orbital wavefunction at wavevector $\mathbf{k}$. 
$\Delta V_{KS}^{\mathbf{q}\lambda}$ is the response of the self-consistent Kohn-Sham potential with respect to the phonon mode $\lambda$ at wavevector $\mathbf{q}$.

Because we use a large supercell that contains a conducting molecule and two electrodes, $\Gamma$-point sampling can be used safely. 
Thus, by dropping wavevector indices for electrons and vibrations, the interaction Hamiltonian can be written in the simpler form:
\begin{equation}
\mathcal{H}_{el-vib} = \sum_{\lambda} \sum_{mn} g_{\lambda}^{mn} a^{\dagger}_{m} a_{n} \left(b^{\dagger}_{\lambda} + b_{\lambda} \right),
\end{equation}
where
\begin{equation}
g_{\lambda}^{mn} = \left(\frac{\hbar}{2\omega_{\lambda}} \right)^{1/2} 
\langle \psi_{m}| \Delta v_{KS}^{\lambda} | \psi_{n}\rangle.
\end{equation}

\subsection{Maximally Localized Wannier Functions}
As discussed in the introduction, DFT calculations based on a planewave basis can provide a very accurate description on the electronic structure of the system, in particular 
in comparison with a localized basis set. However, since Bloch orbitals are intrinsically delocalized, they are not suitable to quantum transport calculation based on 
Green's functions, in which spatial seperation between electrodes and the conductor is required in the Hamiltonian description. 
In this work maximally localized Wannier functions (MLWFs)\cite{1997Marzari,2002Souza} are used 
in order to transform Bloch wavefunctions into localized functions. When there are $N$ isolated bands, WFs can be defined as
\begin{equation}
\label{Wannier_rotation}
|\omega_{n\mathbf{R}}\rangle = \sum_{\mathbf{k}}\sum_{m}^{N} e^{-i\mathbf{k}\cdot \mathbf{R}} U^{(\mathbf{k})}_{mn} |\psi_{m\mathbf{k}}\rangle, 
\end{equation}
where $\mathbf{R}$ is the Bravais lattice vector and the Wannier rotation matrix $U^{(\mathbf{k})}$ is unitary. In the MLWF construction, $U^{(\mathbf{k})}$ is determined by 
minimizing the mean square spread of the resulting Wannier functions, defined as 
\begin{eqnarray}
\Omega &=& \sum_{n}^{N} \left[\langle r^2 \rangle_n - \langle \mathbf{r} \rangle_{n}^{2} \right]\nonumber\\
&=& \sum_{n}^{N} \left[\langle \omega_{n\mathbf{0}} | r^2 | \omega_{n\mathbf{0}} \rangle - \langle \omega_{n\mathbf{0}} | \mathbf{r} | \omega_{n\mathbf{0}} \rangle^2  \right].
\end{eqnarray}
Note that by construction MLWFs are strongly localized in real space. 
Since the Wannier transformation is an exact unitary mapping, the numerical accuracy of planewave-based DFT calculations is preserved in the transformation. 

For a metallic system where all energy bands are connected together, one needs to perform a \emph{disentanglement}\cite{2002Souza} 
in which a maximally-connected $N$-dimensional subspace is extracted from the entire entangled manifold:
\begin{equation}
\label{disentanglement}
|\psi^{opt}_{n\mathbf{k}}\rangle = \sum_{m \in N^{(k)}_{win}} U^{dis(k)}_{mn} |\psi_{m\mathbf{k}}\rangle;
\end{equation}
here, $N^{(k)}_{win}$ is the number of entangled bands in a desired energy window at a wavevector $\mathbf{k}$. 
For details, we refer to Ref. \cite{2002Souza}. 

Using Eq.(\ref{Wannier_rotation}) and Eq.(\ref{disentanglement}), one can readily obtain the electronic Hamiltonian and electron-vibration interaction Hamiltonian in the MLWF basis. 
In particular, we provide an explicit transformation of Hamiltonian from the Bloch representation to Wannier one for the $\Gamma$-point sampling case, which is used in  
large supercell calculations. Under the Wannier transformation, 
\begin{equation}
|\omega_{n}\rangle = \sum_{i,j} U_{in} U^{dis}_{ji} |\psi_{j}\rangle = \sum_{j} \left(U^{dis} U\right)_{jn} |\psi_{j}\rangle,
\end{equation}
the electronic Hamiltonian and electron-vibration interactions read
\begin{eqnarray}
\mathcal{H}_{e} &=& \sum_{m,n} \mathcal{H}_{mn} c_{m}^{\dagger} c_{n} \\
\mathcal{H}_{el-vib} &=& \sum_{\lambda} \sum_{m,n} \mathcal{M}^{\lambda}_{mn} c_{m}^{\dagger} c_{n} \left(b^{\dagger}_{\lambda} + b_{\lambda} \right),
\end{eqnarray}
where
\begin{eqnarray}
\mathcal{H}_{mn} &=& \langle\omega_{m} | \mathcal{H}_{e} | \omega_{n} \rangle \\
                 &=& \sum_{ij} \left(U^{dis}U \right)^{*}_{im} \langle \psi_{i} | \mathcal{H}_{e} | \psi_{j} \rangle \left(U^{dis}U \right)_{jn} \\
\mathcal{M}^{\lambda}_{mn} &=& \sqrt{\frac{\hbar}{2\omega_{\lambda}}} \langle\omega_{m} | \Delta V^{\lambda} | \omega_{n} \rangle \\
                 &=& \sqrt{\frac{\hbar}{2\omega_{\lambda}}} \sum_{ij} \left(U^{dis}U \right)^{*}_{im} \langle \psi_{i} | \Delta V^{\lambda} | \psi_{j} \rangle \left(U^{dis}U \right)_{jn}.
\end{eqnarray}
Here $c_{m}^{\dagger}(c_{m})$ are the electron creation (annihilation) operators for Wannier function $|\omega_{m}\rangle$.

\section{Quantum Transport Theory}
\label{section3}

\subsection{System Division}

\begin{figure}[b]
\begin{center}
\includegraphics[width=0.4\textwidth, clip=true]{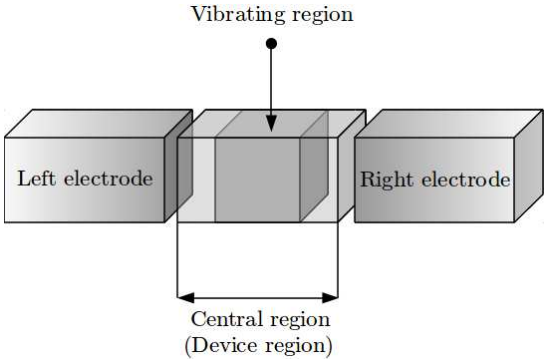} 
\end{center}
\caption{\label{lcr.geometry}Two-terminal geometry used in quantum transport calculation. The device region 
consists of a vibrating molecule and some of the bulk electrode relaxed surface layers.}
\end{figure}

A two-terminal system can be divided into three parts: the left
electrode (L), the central region (C), and the right electrode (R).
The central region is defined as the region that contains the molecules,
defects, or nanoscale conductors of interest. With a localized orbitals basis the corresponding electronic Hamiltonian has the following structure, 
\begin{equation}
\mathcal{H}_{e}=\left(\begin{array}{ccc}
\mathcal{H}_{L} & \mathcal{H}_{LC} & 0\\
\mathcal{H}_{CL} & \mathcal{H}_{C} & \mathcal{H}_{CR}\\
0 & \mathcal{H}_{RC} & \mathcal{H}_{R}\end{array}\right),\label{eq:lcr_hamiltonian}
\end{equation}
where $\mathcal{H}_{CL}=\mathcal{H}_{LC}^{\dagger}$ and $\mathcal{H}_{CR}=\mathcal{H}_{RC}^{\dagger}$.
$\mathcal{H}_{L(R)C}$ is the coupling matrix between the left (right) electrode and the central region. 
Note that the central region does not necessarily include only the conducting molecule, but, as noted in Eq. (\ref{eq:lcr_hamiltonian}),
should be large enough to make the direct coupling between the electrodes zero, $\mathcal{H}_{LR}=\mathcal{H}_{RL}=0$. 
This can be achieved by including some surface atoms of the electrodes into the central region and defining them as an \emph{extended molecule}.
For the elastic quantum conductance calculation this condition is sufficient to select the device region. However, in DFT-NEGF it is assumed that
the conducting electrons are scattered by the molecular vibrations only in the device region, while the electrons in the electrodes are treated as noninteracting quasi-particles. Therefore
the central region should be large enough to guarantee that the electron-vibration interactions converge to zero outside it.

Once the electronic Hamiltonian $\mathcal{H}_{e}$ has been obtained, the \emph{non-interacting} retarded Green's function $G^{r}_{0}(\varepsilon)$ for the conductor region can be derived as follows:
\begin{equation}
\label{eq:0th_RGF}
G_{0}^{r}=\left[\left(\varepsilon+i\eta\right)\mathcal{I}-\mathcal{H}_{C}-\Sigma_{L}^{r}-\Sigma_{R}^{r}\right]^{-1},
\end{equation}
where \begin{eqnarray}
\label{eq:left_self}
\Sigma_{L}^{r} & = & \mathcal{H}_{CL}\frac{1}{\left(\varepsilon+i\eta\right)\mathcal{I}_{L}-\mathcal{H}_{L}}\mathcal{H}_{LC}\\
\label{eq:right_self}
\Sigma_{R}^{r} & = & \mathcal{H}_{CR}\frac{1}{\left(\varepsilon+i\eta\right)\mathcal{I}_{R}-\mathcal{H}_{R}}\mathcal{H}_{RC}.
\end{eqnarray}
Here, $\Sigma_{L/R}^{r}$ are the lead self-energies, which describe the effects of the coupling between the conductor and the leads,  $\eta$ is an infinitesimal positive number, and $\mathcal{I}$ the identity matrix. 
In case of a non-orthogonal atomic basis set calculation, $\mathcal{I}$ should be replaced by the overlap matrix. However, since MLWFs are orthonormal by construction, we can safely use the identity. 
Up to now electron-vibration interactions have not been included in the Green's function.

Since $\mathcal{H}_{L/R}$ are semi-infinite it would be impossible to calculate directly $\left[\left(\varepsilon+i\eta\right)\mathcal{I}_{L/R}-\mathcal{H}_{L/R}\right]^{-1}$.
However, we can use the fact that the localized orbitals of the central region have nonzero overlap only in a finite range close to the conductor. 
If $\mathcal{P}_{L(R)}$ denotes a projection operator onto the subspace of the left (right) electrode that has finite coupling with the central region, 
$\mathcal{H}_{LC(RC)} = \mathcal{P}_{L(R)}\mathcal{H}_{LC(RC)}$ and $\mathcal{H}_{CL(CR)} = \mathcal{H}_{CL(CR)}\mathcal{P}_{L(R)}$.
Therefore, what we need to calculate is 
$\mathcal{P}_{L(R)}\left[\left(\varepsilon+i\eta\right)\mathcal{I}_{L(R)}-\mathcal{H}_{L(R)}\right]^{-1}\mathcal{P}_{L(R)}$,
which is the surface Green's function. This can be obtained by the efficient and fast iterative method first suggested by Sancho et al\cite{Sancho1,Sancho2,Sancho3}.

\subsection{Interacting Green's Functions}

The full Green's function with the electron-vibration coupling can be obtained from diagrammatic perturbation theory based on non-equilibrium Green's functions, known as the Keldysh formalism.
From the theory one can obtain four kind of interacting Green's functions: $G^{r,a}(\varepsilon)$ and $G^{\lessgtr}(\varepsilon)$.
They satisfy the Dyson equation and Keldysh equation respectively, 
\begin{eqnarray}
\label{eq:dyson_eq}
G^{r/a} & = & G_{0}^{r/a}+G_{0}^{r/a}\Sigma_{vib}^{r/a}G^{r/a}\\
\label{eq:Keldysh_eq}
G^{\lessgtr} & = & G^{r}\left(\Sigma_{L}^{\lessgtr}+\Sigma_{R}^{\lessgtr}+\Sigma_{vib}^{\lessgtr}\right)G^{a}.
\end{eqnarray}
The lead lesser and greater self-energies $\Sigma_{\alpha}^{<}$ and $\Sigma_{\alpha}^{>}$ are 
\begin{eqnarray}
\Sigma_{\alpha}^{<}(\varepsilon) & = & i\Gamma_{\alpha}(\varepsilon)f_{\alpha}(\varepsilon)\\
\Sigma_{\alpha}^{>}(\varepsilon) & = & i\Gamma_{\alpha}(\varepsilon)\left(f_{\alpha}(\varepsilon)-1\right),
\end{eqnarray}
where $\Gamma_{\alpha}\equiv i\left(\Sigma_{\alpha}^{>}-\Sigma_{\alpha}^{<}\right)$
and $f_{\alpha}$ is the Fermi-Dirac distribution with chemical potential $\mu_{\alpha}$. In principle the electron-vibration self-energy 
$\Sigma_{vib}^{r/a}$ and $\Sigma_{vib}^{\lessgtr}$ contain all possible diagrams that satisfy Feynman rules. In this work the electron-vibration
self-energies $\Sigma_{vib}^{r/a}$ and $\Sigma_{vib}^{\lessgtr}$ are calculated from the lowest order diagrams in the diagrammatic expansion. 
The lesser and greater electron-vibration self-energies $\Sigma_{vib}^{\lessgtr}(\varepsilon)$ are
\begin{eqnarray}
\Sigma_{vib}^{\lessgtr}(\varepsilon) &=& \sum_{\lambda}\Sigma_{vib,\lambda}^{\lessgtr}\nonumber\\
&=&\label{eq:elph-lessgtr-SCBA}
\sum_{\lambda}i\int_{-\infty}^{\infty}\frac{d\varepsilon^{\prime}}{2\pi}D_{\lambda}^{\lessgtr}\left(\varepsilon-\varepsilon^{\prime}\right)\mathcal{M}_{\lambda}G^{\lessgtr}\left(\varepsilon^{\prime}\right)\mathcal{M}_{\lambda},
\end{eqnarray}
where $D_{\lambda}^{\lessgtr}(\varepsilon)$ are the zeroth-order unperturbed correlation functions for the $\lambda$th vibrational mode,
\begin{equation}
\label{eq:vib-GF}
D_{\lambda}^{\lessgtr}(\varepsilon)=-2\pi i\left[N_{\lambda}\delta\left(\varepsilon\mp\omega_{\lambda}\right)+\left(N_{\lambda}+1\right)\delta\left(\varepsilon\pm\omega_{\lambda}\right)\right].
\end{equation}
Here $N_{\lambda}$ is the vibronic occupation number. 
From the useful relation between the self-energies $\Sigma^{r}-\Sigma^{a}=\Sigma^{>}-\Sigma^{<}$ and the Kramers-Kronig relation one can readily obtain 
\begin{equation}
\label{eq:elph-retarded-SCBA}
\Sigma_{vib}^{r}(\varepsilon)=\frac{1}{2}\left(\Sigma_{vib}^{>}-\Sigma_{vib}^{<}\right)-\frac{i}{2}H\left[\Sigma_{vib}^{>}-\Sigma_{vib}^{<}\right],
\end{equation}
where $H[\cdots]$ is the Hilbert transformation. 
Equations (\ref{eq:elph-lessgtr-SCBA}) and (\ref{eq:elph-retarded-SCBA}) together with Eq. (\ref{eq:dyson_eq}) and (\ref{eq:Keldysh_eq}) constitute a set of coupled equations. 
These equations can be solved in the self-consistent Born approximation (SCBA)\cite{2004Frederiksen,2004Pecchia,2007Frederiksen}. 
However, the SCBA is computationally very demanding, and in case of weak electron-vibration coupling, the lowest-order perturbation theory (LOPT)
may be a good approximation. In this approximation, the full correlation function $G^{\lessgtr}$ in Eq.(\ref{eq:elph-lessgtr-SCBA}) is replaced by the non-interacting $G^{\lessgtr}_0$:
\begin{equation}
\label{eq:elph-lessgtr-LOPT}
\Sigma_{vib}^{\lessgtr,(2)}(\varepsilon) = \sum_{\lambda}i\int_{-\infty}^{\infty}\frac{d\varepsilon^{\prime}}{2\pi}D_{\lambda}^{\lessgtr}\left(\varepsilon-\varepsilon^{\prime}\right)\mathcal{M}_{\lambda}G^{\lessgtr}_{0}\left(\varepsilon^{\prime}\right)\mathcal{M}_{\lambda},
\end{equation}
\begin{equation}
\label{eq:elph-retarded-LOPT}
\Sigma_{vib}^{r,(2)}(\varepsilon)=\frac{1}{2}\left(\Sigma_{vib}^{>,(2)}-\Sigma_{vib}^{<,(2)}\right)-\frac{i}{2}H\left[\Sigma_{vib}^{>,(2)}-\Sigma_{vib}^{<,(2)}\right].
\end{equation}
Within the same order of electron-vibration coupling $\mathcal{M}_{\lambda}$, the Green's functions are calculated as follows:
\begin{eqnarray}
\label{eq:DysonEq-LOPT}
G^{r/a} &\approx& G_{0}^{r/a} + G_{(2)}^{r/a}\nonumber\\
&=& G_{0}^{r/a}+G_{0}^{r/a}\Sigma_{vib}^{r/a,(2)}G_{0}^{r/a}\\
G^{\lessgtr} &\approx& G^{r}_{0} \left( \Sigma_{L}^{\lessgtr}+\Sigma_{R}^{\lessgtr} \right) G_{0}^{a} \nonumber \\
&&+ G^{r}_{(2)} \left( \Sigma_{L}^{\lessgtr}+\Sigma_{R}^{\lessgtr} \right) G_{0}^{a} + G^{r}_{0} \left( \Sigma_{L}^{\lessgtr}+\Sigma_{R}^{\lessgtr} \right) G^{a}_{(2)}\nonumber \\
\label{eq:KeldyshEq-LOPT} 
&&+ G^{r}_{0} \Sigma_{vib}^{\lessgtr,(2)} G^{a}_{0},
\end{eqnarray}
where $G_{(2)}^{r/a} \equiv G_{0}^{r/a}\Sigma_{vib}^{r/a,(2)}G_{0}^{r/a}$.

\subsection{Current Calculations}
Once the full interacting Green's functions and self-energies are calculated, physical quantities of interest can be obtained. 
In order to calculate transport properties, we adopt the Meir-Wingreen transport formula, which is widely applied to mesosopic and nanoscale transport problems\cite{MeirWingreen1}. 
The net electric current $I_{\alpha}$ entering the electrode is 
\begin{eqnarray}
I_{\alpha} &=& e\left\langle \frac{d}{dt}\hat{N}_{\alpha}\right\rangle \nonumber\\
&=&\label{eq:MeirWingreen}
\frac{2e}{\hbar}\int_{-\infty}^{\infty}\frac{d\varepsilon}{2\pi}\textrm{Tr}\left[\Sigma_{\alpha}^{>}(\varepsilon)G^{<}(\varepsilon)-\Sigma_{\alpha}^{<}(\varepsilon)G^{>}(\varepsilon)\right],
\end{eqnarray}
where $\hat{N}_{\alpha}$ is the electron number operator of the $\alpha$ electrode. 
The correlation function $G^{\lessgtr}$ in Eq.(\ref{eq:MeirWingreen}) is calculated in the SCBA and the LOPT, as introduced in the previous section. 
It should be noted that these approximations for the Green's functions satisfy the current conservation condition, $\sum_{\alpha} I_{\alpha} = 0$. 
It is tempting to solve Eq.(\ref{eq:dyson_eq}) and (\ref{eq:Keldysh_eq}) together with Eq.(\ref{eq:elph-lessgtr-LOPT}) and Eq.(\ref{eq:elph-retarded-LOPT}), an approach known as
the first Born approximation (1BA). Although this approximation is much easier to calculate than the SCBA, the 1BA does not generally guarantee current conservation. 

Similarly, one can calculate the net energy current collected in the electrode,
\begin{equation}
P_{\alpha} = \frac{2}{\hbar}\int_{-\infty}^{\infty}\frac{d\varepsilon}{2\pi} \varepsilon \textrm{Tr}\left[\Sigma_{\alpha}^{>}(\varepsilon)G^{<}(\varepsilon)-\Sigma_{\alpha}^{<}(\varepsilon)G^{>}(\varepsilon)\right].
\end{equation}
According to energy conservation, the total sum of the energy current to the electrodes and the local vibrations should be zero, i.e. $P_{L} + P_{R} + P_{vib}=0$. 
From this condition, the net power transferred to local molecular vibrations is given by 
\begin{eqnarray}
P_{vib} &=& \sum_{\lambda} P_{\lambda} \nonumber\\
        &=& \sum_{\lambda} \frac{2}{\hbar}\int_{-\infty}^{\infty}\frac{d\varepsilon}{2\pi} \varepsilon \textrm{Tr}\left[\Sigma_{vib,\lambda}^{>}(\varepsilon)G^{<}(\varepsilon)-\Sigma_{vib,\lambda}^{<}(\varepsilon)G^{>}(\varepsilon)\right].
\end{eqnarray}

\subsection{Vibrational Population}
Due to electron-vibration interactions, conducting electrons can exchange their energy with molecular vibrations by absorbing or emitting local vibrational quanta. 
Therefore the vibrational population $N_{\lambda}$ in Eq.(\ref{eq:vib-GF}) does not in general follow a Bose-Einstein distribution. 
In order to describe a non-equilibrium vibrational population, a semiclassical rate equation has been suggested\cite{2007Frederiksen,2007Pecchia},
\begin{equation}
\label{RateEq1}
\frac{d}{dt}N_{\lambda}  =  \frac{P_{\lambda}}{\hbar\omega_{\lambda}}-\gamma_{\lambda}\left(N_{\lambda}-n_{B}\left(\hbar\omega_{\lambda}\right)\right).
\end{equation}
The first term on the right hand side represents the net molecular vibrations excited by the conducting electrons. 
The second term describes a heat dissipation process that makes local vibrations equilibrated to the Bose-Einstein distribution at the electrode temperature. 
Molecular vibrations are not isolated, but are mechanically coupled to the surrounding (e.g. the bulk phonons of the electrodes). Due to this coupling, 
local vibrations in the molecule decay into bulk phonons, and $\gamma_{\lambda}$ represents this decay rate.
Note that here we restrict ourselves to consider only the \emph{harmonic} coupling to bulk phonons. 
Essentially the harmonic coupling describes \emph{one local vibration to one phonon} transitions, which mean that a local vibration decays by exciting one phonon mode in the electrodes. 
Following Ref.(\cite{2007Pecchia}), the first term in Eq.(\ref{RateEq1}) can be decomposed into absorption and emission processes,
\begin{equation}
\label{RateEq2}
\frac{d}{dt}N_{\lambda}  = \left(N_{\lambda}+1\right)E_{\lambda}-N_{\lambda}A_{\lambda} - \gamma_{\lambda}\left(N_{\lambda}-n_{B}\left(\hbar\omega_{\lambda}\right)\right).
\end{equation}
 Here the absorption and emission rates are expressed as follows:
\begin{eqnarray}
\label{eq:absorption_rate}A_{\lambda} &=& \frac{2}{\hbar} \int_{-\infty}^{\infty} \frac{d\varepsilon}{2\pi} \textrm{Tr} \left[\mathcal{M}^{\lambda} G^{<}(\varepsilon-\hbar\omega_{\lambda}) \mathcal{M}^{\lambda} G^{>}(\varepsilon) \right]\\
\label{eq:emission_rate}E_{\lambda} &=& \frac{2}{\hbar} \int_{-\infty}^{\infty} \frac{d\varepsilon}{2\pi} \textrm{Tr} \left[\mathcal{M}^{\lambda} G^{<}(\varepsilon+\hbar\omega_{\lambda}) \mathcal{M}^{\lambda} G^{>}(\varepsilon) \right].
\end{eqnarray}
The steady state solution for the vibrational populations can be immediately obtained from Eq.(\ref{RateEq2}),
\begin{equation}
\label{steady_population}
N_{\lambda} = \frac{n_{B}\left(\hbar\omega_{\lambda}\right) \gamma_{\lambda} + E_{\lambda}}{A_{\lambda} + \gamma_{\lambda} - E_{\lambda}}.
\end{equation}
Note that this steady state solution exists only when $A_{\lambda} + \gamma_{\lambda} > E_{\lambda}$. 
Otherwise, Eq.(\ref{RateEq2}) gives an exponentially increasing population as a function of time, which implies a vibrational instability. 

In order to calculate vibrational decay rates, let us consider a generic case where localized vibrations are harmonically coupled to bulk phonons (heat reservoir).
Localized molecular vibrations and bulk phonons are described by collections of harmonic oscillators in the second quantized form:
\begin{eqnarray}
H_{S} & = & \sum_{\lambda} \hbar \omega_{\lambda} \left(b^{\dagger}_{\lambda} b_{\lambda} + \frac{1}{2} \right)\\
H_{B} & = & \sum_{k} \hbar \bar{\omega}_{k} \left(\bar{b}^{\dagger}_{k} \bar{b}_{k} + \frac{1}{2} \right), 
\end{eqnarray}
where $b_{\lambda}(b^{\dagger}_{\lambda})$ and $\bar{b}_{k}(\bar{b}^{\dagger}_{k})$ are annihilation (creation) operators for localized vibrations and bulk phonons respectively.
When the $\lambda$th molecular vibrational state is occupied, the decay rate at which the vibration disappears by exciting bulk phonons can be calculated by using \emph{Fermi golden} rule\cite{2006Andrianov, 2008Sakong, 2010Sakong},
\begin{equation}
\label{eq:decaying_rate_FGR}
\Gamma_{\lambda\rightarrow\textrm{bath}} = \frac{2\pi}{\hbar} \sum_{k\in \textrm{bath}} \left|_{B}\langle k | \mathcal{H}_{C} | \lambda \rangle_{S}  \right|^2 \delta\left(\hbar \bar{\omega}_{k} - \hbar \omega_{i} \right),
\end{equation}
where $|\lambda\rangle_{S} = b^{\dagger}_{\lambda} |0\rangle$ and $|k\rangle_{B} = \bar{b}^{\dagger}_{k} |0\rangle$. 
Here $H_{C}$ denotes the harmonic coupling between local vibrations and the heat reservoir.
In fact, after some lengthy algebra, one can show that Eq.(\ref{eq:decaying_rate_FGR}) is equivalent to the following expression:
\begin{equation}
\Gamma_{\lambda\rightarrow\textrm{bath}} = -\frac{1}{\omega_{\lambda}} \vec{u}_{\lambda}^{T} \textrm{Im} \Pi^{r} \left(\omega_{\lambda} \right) \vec{u}_{\lambda},
\end{equation}
where $\vec{u}_{\lambda}$ is the normal mode vector of the $\lambda$th local vibration\cite{2007Romano,2009Engelund}. 
$\Pi^r(\omega)$, which is defined as the retarded \emph{heat bath self-energy}, 
is the \emph{mechanical} counterpart to the electronic lead self-energy $\Sigma^r(\varepsilon)$. 
The numerical methods used to calculate $\Sigma^r(\varepsilon)$\cite{Sancho1,Sancho2,Sancho3} can be directly applied to the calculation of $\Pi^r$.

\section{Applications}
\label{section4}
In this section we present two applications of our \emph{ab initio} inelastic transport approach.
As a first application we calculate inelastic transport properties of a benzene molecule connected to 
monoatomic carbon chains (cumulenes). For this benchmark system we apply both the SCBA and the LOPT for equilibrium and non-equilibrium solutions. 
As a second example we replace cumulene with $(3,3)$ single-wall carbon nanotube (CNT). 
In this molecular junction we focus on the calculation of realistic decay rates and non-equilibrium vibration populations. 
Technically, all DFT calculations are performed using the Perdew-Zunger local density approximation\cite{1981Perdew}, norm-conserving
pseudopotentials\cite{1991Troullier}, and a plane-wave basis with a cutoff of $55 \textrm{ Ry}$. 

\subsection{Cumulene-Benzene-Cumulene junction}
\begin{figure}[!t]
\begin{center}
\includegraphics[width=0.45\textwidth, clip=true]{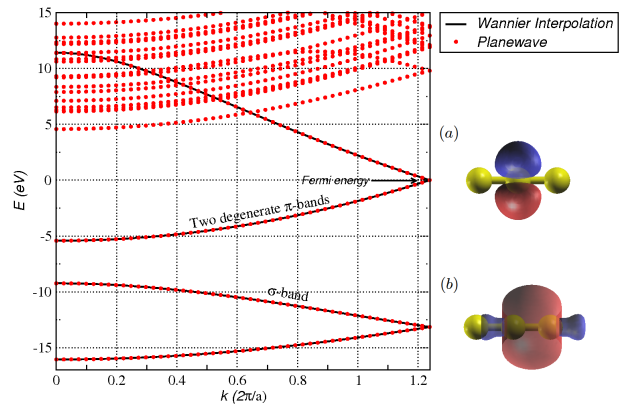} 
\end{center}
\caption{\label{cumulene.bands.wfs}(Color online) Band structure of cumulene for a two-atom unit cell.
(Red dots: direct DFT calculation; black solid lines: Wannier interploted bands). (a) $p$-type Wannier function at
an atomic site (b) $\sigma$-like Wannier function at a mid-bond site. }
\end{figure}

\begin{figure}[t]
\begin{center}
\includegraphics[width=0.45\textwidth, clip=true]{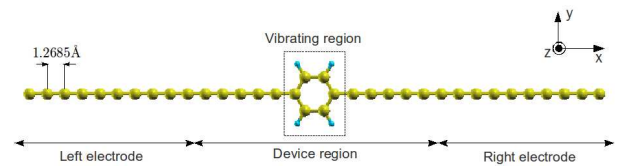} 
\end{center}
\caption{\label{cumulene-C6H4-cumulene}(Color online) Cumulene-benzene-cumulene junction.}
\end{figure}

We first study a benzene molecule connected to a monoatomic carbon chain. 
It is known that cumulene is subject to a Peierls distortion: It readily becomes dimerized and opens an energy band gap favored by a lower energy structure. 
We thus freeze the structure of cumulene and use it as a metallic electrode. 
We construct two $p$-orbitals and one $\sigma$-like mid-bond Wannier functions (see Fig.\ref{cumulene.bands.wfs}) for this electrode. 
The two $p$-orbitals, $p_{y}$ and $p_{z}$, are perpendicular to the transport direction, which is along the $x$-axis in our calculation. 
Figure \ref{cumulene.bands.wfs} shows the band structure of cumulene. The energy bands are obtained either from
a direct planewave-based DFT calculation, or by Wannier band interpolation, and are in excellent agreement.
While the lowest energy band originates from $\sigma$-orbitals, $p$-orbitals give rise to doubly degenerate $\pi$-bands around the Fermi level, and 
transport properties at the Fermi energy are characterized by these two $\pi$-bands.

\begin{figure}[!t]
\begin{center}
\includegraphics[width=0.45\textwidth, clip=true]{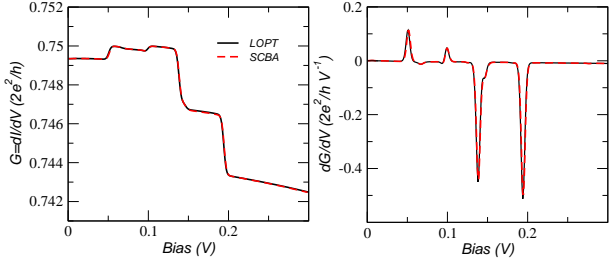} 
\end{center}
\caption{\label{cc-C6H4-eq}(Color online) Differential conductance $G=dI/dV$ and its derivative $dG/dV$ with equilibrum vibrationl populations
calculated using LOPT (black solid line) and SCBA (red dashed line). 
At lower bias two differential conductance increases are observed. At higher bias, two large conductance drops
occur. These conductance changes correspond to peaks in $dG/dV$. }
\end{figure}

Figure \ref{cumulene-C6H4-cumulene} shows the supercell geometry used in the transport calculations. 
The benzene molecule alone is allowed to vibrate.
The device region, containing the vibrating region and part of the cumulene, is taken to be large enough to make sure that there is no direct coupling between electrodes,
and that the electron-vibration coupling is zero outside the device region. 
For the benzene molecule, $p_{z}$-type Wannier functions on carbon atoms and $\sigma$-like Wannier functions on $\textrm{C}-\textrm{C}$ and $\textrm{C}-\textrm{H}$ bonds are constructed.

\begin{figure}[t]
\begin{center}
\includegraphics[width=0.45\textwidth, clip=true]{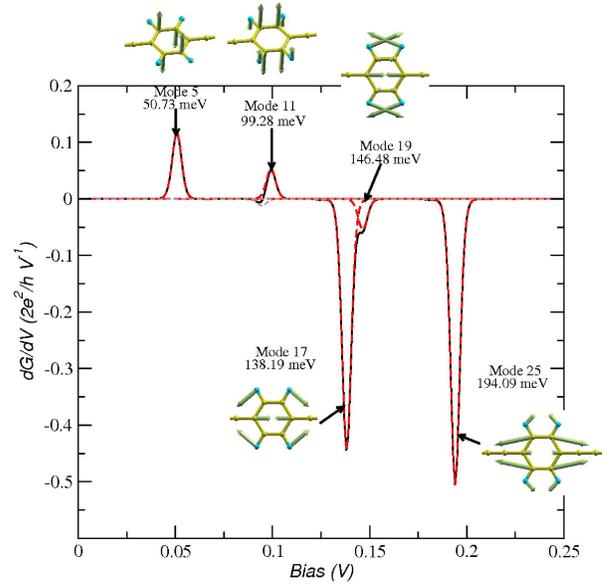} 
\end{center}
\caption{\label{cc-C6H4-eq-modewise-temp3}(Color online) $dG/dV$ in modewise calculation. Five active vibrational modes are found.
The corresponding vibrational configuration are illustrated. While the first two active modes leading to conductance jumps
are out-of-plane motions, the three conductance-drop modes correspond to in-plane vibrations.}
\end{figure}

The differential conductance $G=dI/dV$ and its derivative $dG/dV$ are calculated using either the SCBA or the LOPT scheme. Temperature is taken to be $k_{B}T=1\textrm{meV}$. 
As seen in Fig.\ref{cc-C6H4-eq}, these two approximations display essentially perfect agreement.
Four conductance changes are observed in the differential conductance curve of Fig.\ref{cc-C6H4-eq}. 
The corresponding inelastic transport signals due to electron-vibration interactions appear as peaks in the $d^2 I / dV^2$ plot. 
The peak position on the bias axis corresponds to the vibrational energy involved in the electron-vibration scattering events. 
From four peaks observed in Fig.\ref{cc-C6H4-eq} one might conclude that there are four active vibrational modes, but there is a shoulder on the right side of the third peak. 
This may indicate that there is a fifth active vibrational mode. To investigate which vibrational modes participate in inelastic transport,
we performed \emph{modewise} calculation by keeping only one particular vibrational mode. For clarity the elastic contribution is excluded in the modewise calculation. 
These calculations show that there are five major peaks in $dG/dV$ (Fig. \ref{cc-C6H4-eq-modewise-temp3}).
The corresponding vibrational configurations of these five active modes are also shown in Fig.\ref{cc-C6H4-eq-modewise-temp3}:
While two upward peaks are out-of-plane vibrations, three downward peaks correspond to in-plane motions. 

\begin{figure}[!b]
\begin{center}
\includegraphics[width=0.47\textwidth, clip=true]{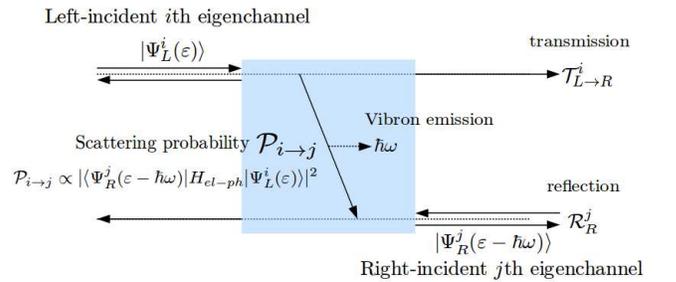} 
\end{center}                                                                                                                                   
\caption{\label{multichannel}Schematic representation of inelastic scattering in the presence of electron-vibration interactions. Solid arrows indicate transmission eigenchannels. }
\end{figure}

\begin{figure}[t]
\begin{center}
\includegraphics[width=0.5\textwidth, clip=true]{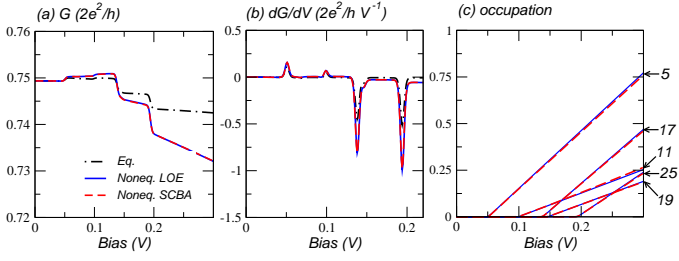} 
\end{center}
\caption{\label{noneq.SCBA.LOE}(Color online)Inelastic transport calculations with nonequilibrium vibrational populations
(blue solid line: LOPT; red dashed line: SCBA; black dot-dashed line: equilibrium case).
(a) differential conductance, (b) second derivative of the current, and (c) vibration populations.}
\end{figure}

Thus, electron-vibration interactions can lead to both differential conductance rises and drops. 
This simultaneous occurrence can be
understood from scattering theory and transmission eigenchannels\cite{1997Brandbyge, 2007Paulsson}, following Ref.\cite{2009Kristensen}. 
As seen in Fig.\ref{multichannel}, let us consider an electron injected from the left electrode on a left-incident $i$th eigenchannel $|\Psi_{L}^{i}\rangle$
at energy $\varepsilon$. If there is no scattering with molecular vibrations, this electron contributes to elastic conductance
\begin{equation}
G_{no\;el-vib} = G_0 \sum_{i} \mathcal{T}^{i}_{L\rightarrow R}(\varepsilon),
\end{equation}
where $G_0=\frac{2e^2}{h}$ and $\mathcal{T}^{i}_{L\rightarrow R}(\varepsilon)$ is the elastic transmission probability of $|\Psi^{i}_{L}(\varepsilon)\rangle$.
Now let us consider that the electron on a left-incident $i$th eigenchannel $|\Psi^{i}_{L}(\varepsilon)\rangle$  is scattered off to a right-incident $j$th eigenchannel
$|\Psi^{j}_{R}(\varepsilon-\hbar\omega)\rangle$ by emitting a vibrational quanta $\hbar\omega$. 
The outgoing state of $|\Psi^{j}_{R}(\varepsilon-\hbar\omega)\rangle$ can go either to the left electrode or to the right one. Since conductance 
measured in the right lead is considered here, the probability to move back to the right electrode for $|\Psi^{j}_{R}(\varepsilon-\hbar\omega)\rangle$
is given by its reflection probability $\mathcal{R}_{R}^{j}(\varepsilon-\hbar\omega)$. 
If $\mathcal{P}_{i \rightarrow j}$ denotes the probability that $|\Psi^{i}_{L}(\varepsilon)\rangle$ is scattered to $|\Psi^{j}_{R}(\varepsilon-\hbar\omega)\rangle$,
then the total conductance can be computed as:
\begin{eqnarray}
G_{el-vib} &=& G_0  \left[ \sum_{i}\left(1- \sum_{j} \mathcal{P}_{i \rightarrow j} \right) \mathcal{T}^i_{L\rightarrow R}(\varepsilon) \right. \nonumber \\
           & & \left. + \sum_{i,j} \mathcal{P}_{i \rightarrow j} \mathcal{R}^{j}_{R}((\varepsilon)-\hbar\omega) \right].
\end{eqnarray}
Thus, the conductance change is given by 
\begin{eqnarray}
\Delta G &=& G_{el-vib} - G_{no\;el-vib} \nonumber \\
         &=& \sum_{i} G_0  \left[ \sum_{j} \mathcal{P}_{i \rightarrow j} \left( \mathcal{R}^j_R(\varepsilon - \hbar\omega) - \mathcal{T}^i_{L\rightarrow R}(\varepsilon) \right) \right] \\
         &\approx& \sum_{i} G_0  \left[ \sum_{j} \mathcal{P}_{i \rightarrow j} \left( \mathcal{R}^j_R(\varepsilon_{F}) - \mathcal{T}^i_{L\rightarrow R}(\varepsilon_{F}) \right) \right].
\end{eqnarray}
Let us examine these relations in detail. 
First, transmission and reflection probabilities are approximated by those at Fermi energy $\varepsilon_{F}$ 
(Since  vibrational energies are generally small, 
one might expect that transmission and reflection probabilities would not change significantly over $[\varepsilon_{F}-\hbar\omega, \varepsilon_{F}+\hbar\omega]$).
Now, note that conductance can increase or decrease depending on the relative magnitude of $\mathcal{P}_{i \rightarrow j} \left( \mathcal{R}^j_R(\varepsilon_{F}) - \mathcal{T}^i_{L\rightarrow R}(\varepsilon_{F}) \right)$
, and that while $\left( \mathcal{R}^j_R(\varepsilon_{F}) - \mathcal{T}^i_{L\rightarrow R}(\varepsilon_{F}) \right)$ does not depend on a vibrational configuration, 
$\mathcal{P}_{i \rightarrow j} \propto \left| \langle \Psi_{R}^{j} | \mathcal{H}_{el-vib}^{\lambda} | \Psi_{L}^{i} \rangle \right|^2$
, which can be calculated from Fermi's golden rule\cite{2003Montgomery1,2003Montgomery2}, is determined by the electron-vibration interaction matrix.

For the cumulene-benzene-cumulene system, it is found that there are two transmission eigenchannels: 
the major transmission channel $|\Psi^1_{L,R}\rangle$ with $\mathcal{T}^{1}_{L\rightarrow R}=\mathcal{T}^{1}_{R\rightarrow L}=0.741$ 
and the minor channel $|\Psi^2_{L,R}\rangle$ with $\mathcal{T}^{2}_{L\rightarrow R}=\mathcal{T}^{2}_{R \rightarrow L}=0.004$.
$p_{z}$ orbitals on the cumulene wire and the benzene molecule constitute the major transmission eigenchannel $|\Psi^1_{L,R}\rangle$. $p_{y}$ orbitals
on the cumulene wire and $\sigma$ bonds of the benzene molecule contribute to the minor transmission channel $|\Psi^2_{L,R}\rangle$.
In addition, while $|\Psi^1_{L,R}\rangle$ is symmetric with respect to the $zx$-plane, $|\Psi^2_{L,R}\rangle$ is anti-symmetric. 
In other words, when $\hat{P}_{zx}$ denotes the reflection operator with respect to $zx$-plane, 
$\hat{P}_{zx}|\Psi^1_{L,R}\rangle = |\Psi^1_{L,R}\rangle$ and $\hat{P}_{zx}|\Psi^2_{L,R}\rangle = -|\Psi^2_{L,R}\rangle$ hold.

The first two active vibrational modes $\lambda=5,11$, shown in Fig.\ref{cc-C6H4-eq-modewise-temp3}, which lead to differential conductance jumps, 
show anti-symmetric vibrational motion with respect to the $zx$-plane. 
Then, the corresponding electron-vibration interactions $\mathcal{H}_{el-vib}^{\lambda=5,11}$ satisfy 
\begin{equation}
\label{parity3}
\mathcal{H}_{el-vib}^{\lambda=5,11} = - \hat{P}_{zx} \mathcal{H}_{el-vib}^{\lambda=5,11} \hat{P}_{zx}. 
\end{equation}
Because of these reflection symmetries, 
\begin{equation}
\langle \Psi^{1}_{R} | \mathcal{H}^{\lambda=5,11}_{el-vib} | \Psi^{1}_{L} \rangle = \langle \Psi^{2}_{R} | \mathcal{H}^{\lambda=5,11}_{el-vib} | \Psi^{2}_{L} \rangle = 0,
\end{equation}
i.e. scattering from $|\Psi^{1(2)}_{L}\rangle$ to $|\Psi^{1(2)}_{R}\rangle$ is prohibited. 
Since $\left(\mathcal{R}^{2}_{R} - \mathcal{T}^{1}_ {L\rightarrow R} \right) = \left( \mathcal{R}^{1}_{R} - \mathcal{T}^{2}_ {L\rightarrow R} \right) = 0.255$, 
one obtains a differential conductance rise
\begin{eqnarray}
\label{jump}
\Delta G &=& G_0  \left[ \mathcal{P}_{1 \rightarrow 2}^{\lambda=5,11} (\mathcal{R}^2_{R} - \mathcal{T}^{1}_{L \rightarrow R}) \right. \nonumber \\
         &&  \left. +  \mathcal{P}_{2 \rightarrow 1}^{\lambda=5,11} (\mathcal{R}^1_{R} - \mathcal{T}^{2}_{L \rightarrow R})\right]  > 0.
\end{eqnarray}
In contrast, the last three active modes $\lambda=17,19$ and $25$ are symmetric with respect to the $zx$-plane:
\begin{equation}
\label{parity4}
\mathcal{H}_{el-vib}^{\lambda=17,19.25} = \hat{P}_{zx} \mathcal{H}_{el-vib}^{\lambda=17,19,25} \hat{P}_{zx}. 
\end{equation}
Therefore, one has these reflection selection rules:
\begin{equation}
\langle \Psi^{1}_{R} | \mathcal{H}^{\lambda=17,19,25}_{el-vib} | \Psi^{2}_{L} \rangle = \langle \Psi^{2}_{R} | \mathcal{H}^{\lambda=17,19,25}_{el-vib} | \Psi^{1}_{L} \rangle = 0.
\end{equation}
Since $\mathcal{P}_{1 \rightarrow 1}^{\lambda=17,19,25} \gg \mathcal{P}_{2 \rightarrow 2}^{\lambda=17,19,25} $ from numerical calculation,  
then one finds the three differential conductance drops
\begin{eqnarray}
\label{drop}
\Delta G &=& G_0  \left[ \mathcal{P}_{1 \rightarrow 1}^{\lambda=17,19,25} (\mathcal{R}^1_{R} - \mathcal{T}^{1}_{L \rightarrow R}) \right. \nonumber \\
         &&  \left. +  \mathcal{P}_{2 \rightarrow 2}^{\lambda=17,19,25} (\mathcal{R}^2_{R} - \mathcal{T}^{2}_{L \rightarrow R})\right] < 0,
\end{eqnarray}
where 
$\left(\mathcal{R}^{1}_{R} - \mathcal{T}^{1}_ {L\rightarrow R} \right) = -0.482$  and $\left( \mathcal{R}^{2}_{R} - \mathcal{T}^{2}_ {L\rightarrow R} \right) = 0.992$.
More generally this multichannel analysis shows that differential conductance rises and drops can occur at the same time. 

\begin{figure}[t]
\begin{center}
\includegraphics[width=0.45\textwidth, clip=true]{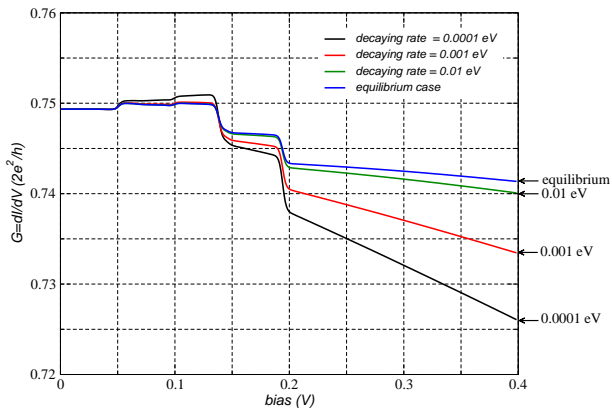} 
\end{center}
\caption{\label{C6H4-noneq-decaying-rate}(Color online)Differential conductance for different decay rates
(black solid line: $\hbar\gamma_\lambda = 0.1 \textrm{ meV}$; red solid line: $\hbar\gamma_\lambda= 1 \textrm{ meV}$;
green solid line: $\hbar\gamma_\lambda = 10 \textrm{ meV}$; blue solid line: equilibrium case ($\hbar\gamma_\lambda \rightarrow \infty$)).}
\end{figure}

Next, let us take into account the effect of nonequilibrium vibrational populations on transport properties.
This situation corresponds to the case where the decay rate of a molecular vibration to its surrounding is 
larger than the emission rate due to electron-vibration scattering. 
In order to compare equilibrium and nonequilibrium vibration cases the decay rate $\hbar\gamma_{\lambda}=0.1\textrm{ meV}$ 
is chosen for all vibrational modes; this condition will be relaxed in the next example. 

As shown in Fig.\ref{noneq.SCBA.LOE} (a), nonequilibrium effects lead to larger slopes in comparison with the equilibrium case. 
Furthermore, the differential conductance change increases at the threshold bias voltage. These two changes
appear as (1) the finite $dG/dV$ value between peaks and (2) increased peak heights in Fig.\ref{noneq.SCBA.LOE} (b). When the
bias exceeds the threshold voltage equal to the vibrational energy, the vibrational population starts to increase, as observed in Fig.\ref{noneq.SCBA.LOE} (c).
This is because of the increase in phase space of conducting electrons that can emit molecular vibration quanta. 
Recalling that electron-vibration scattering is roughly proportional to $N_\lambda$, increased vibration populations enhance
inelastic transport signals in return. Last, we also considered transport calculations where we change the decay rate. 
Figure \ref{C6H4-noneq-decaying-rate} shows that the differential conductance approaches the equilibrium case as the decay rate increases.

\subsection{($3$,$3$) CNT - Benzene - ($3$,$3$) CNT}

\begin{figure}[t]
\begin{center}
\includegraphics[width=0.45\textwidth, clip=true]{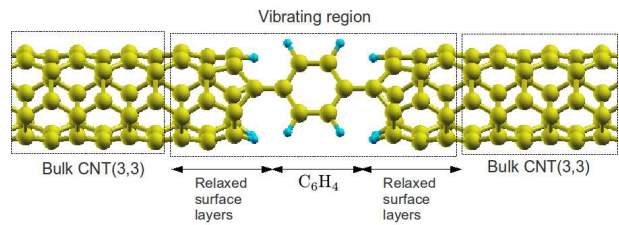}  
\end{center}
\caption{\label{cnt33-C6H4-supercell}(Color online) ($3$,$3$) CNT - Benzene - ($3$,$3$) CNT supercell geometry used in the decay rate calculations.
The vibrating region contains a benzene molecule and three relaxed surface CNT layers.}
\end{figure}

\begin{figure}[t]
\begin{center}
\includegraphics[width=0.45\textwidth, clip=true]{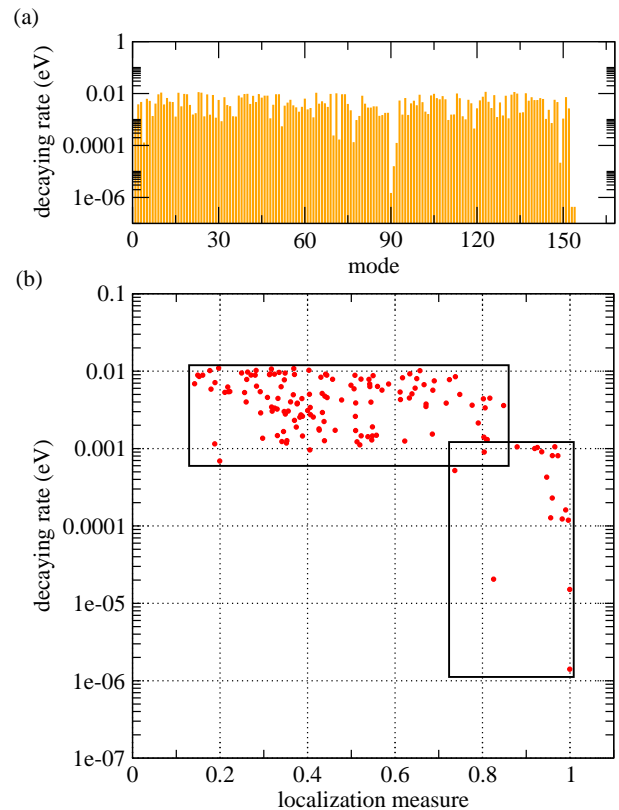} 
\end{center}
\caption{\label{decaying_rate_localization}(Color online) (a) Decay rates for each vibrational mode of the $(3,3)$ CNT-Benzene-$(3,3)$ CNT junction. 
(b) Decay rate vs. localization (see the text for the definition). Decay rates are plotted in a logarithmic scale.}
\end{figure}

\begin{figure}[t]
\begin{center}
\includegraphics[width=0.3\textwidth, clip=true]{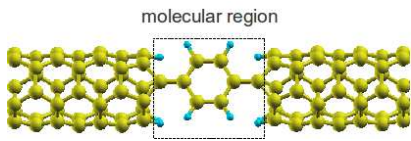}  
\end{center}
\caption{\label{cnt33-C6H4-supercell2}(Color online) Molecular region containing a benzene molecule, anchoring carbon atoms, and hydrogen atoms 
saturating the CNT edge.}
\end{figure}

In the previous benchmark, the cumulene wire, subject to Peierls' instability, was frozen in order to keep its metallic character, and 
the decay rate for molecular vibrations was used as a parameter. 
Here we replace the cumulene wire by a metallic $(3,3)$ carbon nanotube (CNT), which is mechanically stable. 
Using this electrode we derive a fully \emph{ab initio} approach to calculate non-equilibrium populations under electron-vibration interactions. 

Carbon-based nanostructures, such as CNT and graphene, could become new platforms for future nanotechnology applications due to their excellent electronic properties. 
Recently carbon-based nanojunctions have been experimentally fabricated: a pure carbon chain connected to graphene\cite{2009Jin} or CNTs\cite{2010Borrnert}, and organic molecules coupled to 
CNT electrodes with amide linkages\cite{2006Guo}. These experimental achievements have stimulated theoretical and computational studies on carbon-based nanodevices \cite{2007Ke,2009Martins}.
In particular, a benzene molecule connected to CNT electrodes (which is the system of our interest) was suggested as a molecular switch,
operated by controlling the relative angle between $\pi$-orbitals of the benzene and the $\pi$-orbital manifold of CNT electrodes \cite{2009Martins}. 
Functionality and performances of molecular devices are strongly affected by molecular geometries or anchoring points to the electrodes, 
and they may be affected by vibrations induced by conducting electrons. In the worst case, local heating may break down the junctions.

In our work, we choose a vibrating region by defining an extended molecule in which a benzene and the outmost relaxed CNT layers are included. 
This extended molecule is seamlessly connected to the bulk CNT electrodes. The vibrating region contains $56$ atoms in total, and these correspond to $168$ vibrational modes. 
Figure \ref{cnt33-C6H4-supercell} illustrates the supercell geometry used in the decay rate calculations: It contains the extended molecule and two \emph{mechanical}
principal layers\cite{1981Lee} for bulk phonons. 

We proceed as follows. First, by allowing only the extended molecule to vibrate, the electron-vibration interaction $\mathcal{H}_{el-vib}$, 
the vibrational spectrum $\{\hbar\omega_{\lambda} \}$, and the corresponding normal modes are calculated. 
For the decay rates the interatomic force constants for the entire supercell in Fig. \ref{cnt33-C6H4-supercell} are calculated. 
From this interatomic force constants one can extract the harmonic coupling matrix $\mathcal{H}_{C}$ between the extended molecule and the bulk electrodes. 
In addition, we take a periodic unit cell of the bulk $(3,3)$ CNT and calculate its interatomic force constants $H_{B}$. 

The calculated decay rates are shown in Fig.\ref{decaying_rate_localization} (a). 
Note that the decay rates are written in units of $\textrm{eV}$. 
While most of the decay rates are of the order of $10^{-2}$ to $10^{-3}$ $\textrm{eV}$, there are few modes with much smaller rates.  
These small decay rates can arise for two reasons. First, as seen in Fig.\ref{decaying_rate_localization} (a), decay rates start to increase significantly from the $153$th mode on.
These vibrational modes (between $153$ and $168$) have energies higher than the highest $(3,3)$ CNT phonon energy.
Recalling that decay processes based on the harmonic coupling essentially correspond to \emph{one vibration to one phonon}
transitions, there are no bulk phonons to which the vibrational modes lying outside the band width of bulk phonons can transfer 
their vibrational energies. Once the \emph{anharmonic} coupling that makes \emph{one vibration to multi-phonons} transition
possible is taken into account, these modes may have larger decay rates. However, this correction is beyond our work. 

Second, notice that among the vibrational modes whose energies lie inside the bulk phonon dispersions, some of them still have small decay rates.
Most of them correspond to vibrations that are localized inside the benzene molecule, or to \emph{wagging} motions of the surface hydrogen atoms.
Because these motions are spatially well separated from bulk phonons, one may expect that they are less coupled to bulk phonons.
In order to measure how localized these modes are inside the molecule, we define the benzene molecule, two anchoring carbon atoms, and the surface hydrogen atoms as the \emph{molecular region}, as seen in Fig.\ref{cnt33-C6H4-supercell2}.
When $\mathcal{P}_{M}$ denotes a projection operator onto the molecular region, one can find how localized the vibration is inside the molecular region from $|\mathcal{P}_{M} | \lambda\rangle |^2$ where $|\lambda\rangle$ indicates the vibrational state for the normal mode $\lambda$. 
We call $|\mathcal{P}_{M} | \lambda\rangle |^2$ the \emph{localization measure}. 
Figure \ref{decaying_rate_localization} (b) shows a relation between the localization measure and decay rates. 
When the vibration is localized in the molecular region, or equivalently $|\mathcal{P}_{M} |\lambda\rangle|^2$ approaches $1$, its decay rate becomes smaller. 

\begin{figure}[t]
\begin{center}
\includegraphics[width=0.45\textwidth, clip=true]{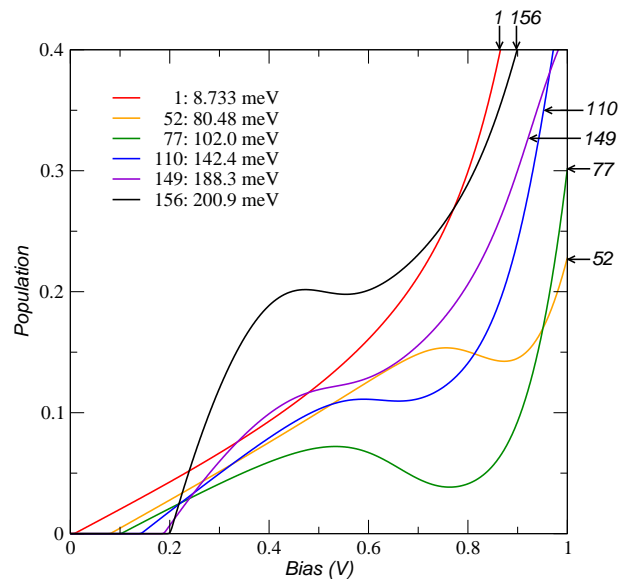}  
\end{center}
\caption{\label{cnt33-C6H4.noneq.phonons}(Color online) Nonequilibrium vibrational populations for the most excitable modes as a function of  a bias voltage. }
\end{figure}

\begin{figure}[t]
\begin{center}
\includegraphics[width=0.45\textwidth, clip=true]{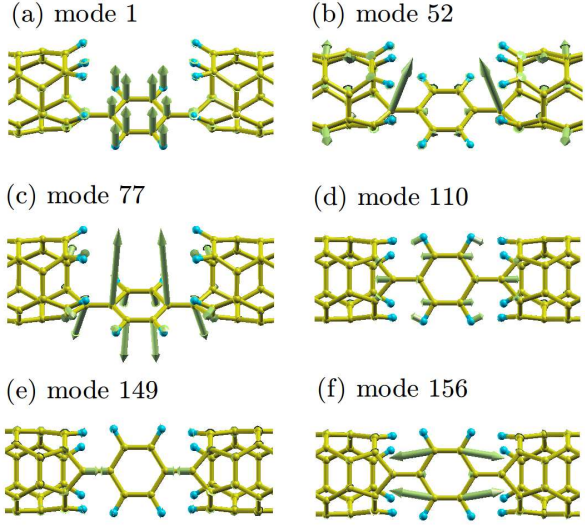}  
\end{center}
\caption{\label{cnt33-C6H4-vib}(Color online) Vibrational configurations for the modes in Fig.\ref{cnt33-C6H4.noneq.phonons}.}
\end{figure}

\begin{figure}[t]
\begin{center}
\includegraphics[width=0.48\textwidth, clip=true]{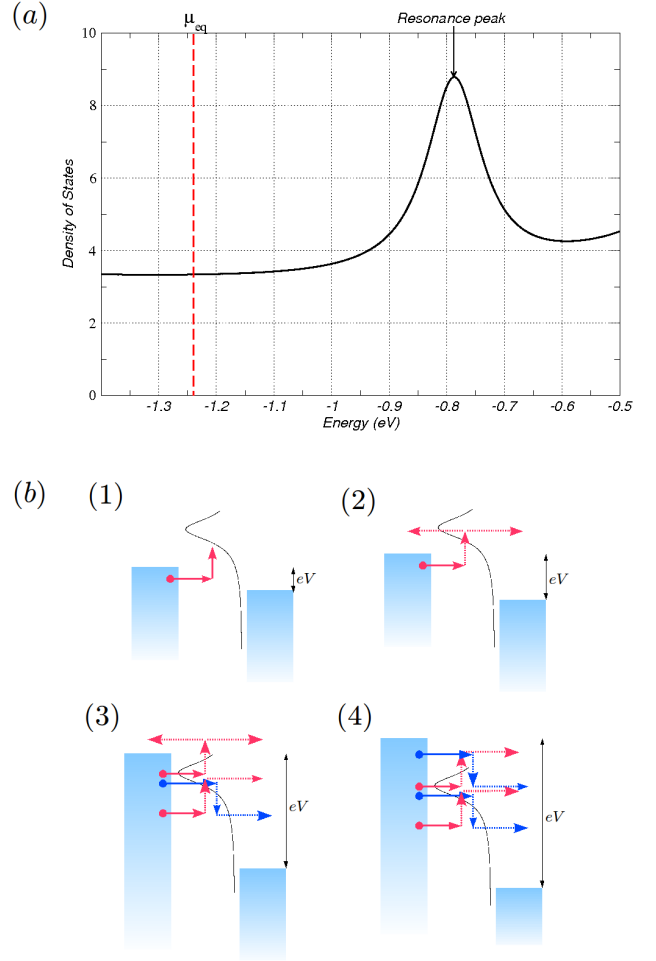}  
\end{center}
\caption{\label{cnt33-C6H4.dos}(Color online) (a) density of states for the device region. Close to the equilibrium Fermi level,
one resonance peak is found. (b) possible absorption (red arrow line) and emission (blue arrow line) processes via the resonant peak as the bias
voltage increases.}
\end{figure}

\begin{figure}[t]
\begin{center}
\includegraphics[width=0.48\textwidth, clip=true]{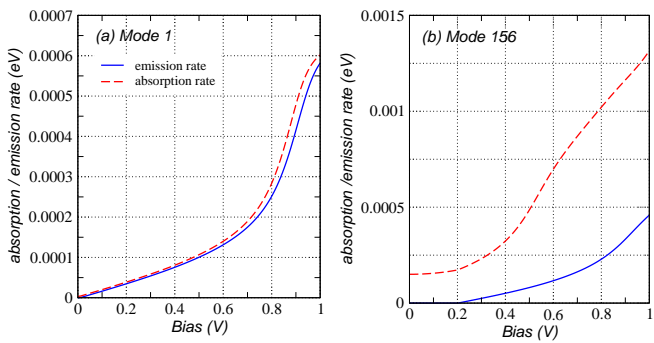}  
\end{center}
\caption{\label{cnt33-C6H4.absorption.emission.rate}(Color online) the absorption (red dashed line)
 and emission (blue solid line) rates for (a) the vibrational mode $1$ (low-energy mode) and (b) $156$ (high-energy mode)}
\end{figure}

\begin{figure}[t]
\begin{center}
\includegraphics[width=0.4\textwidth, clip=true]{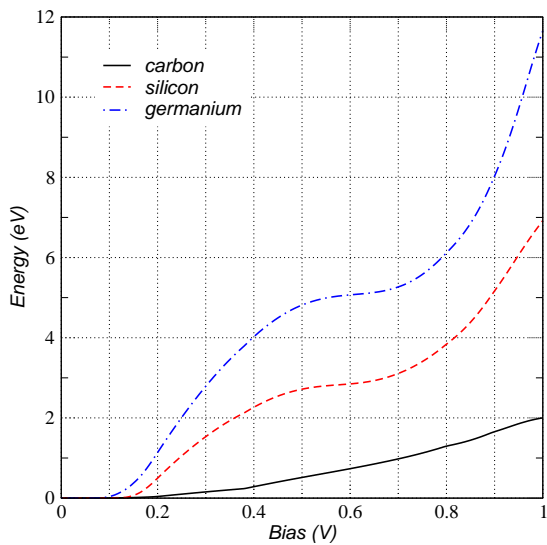}  
\end{center}
\caption{\label{cnt33-C6H4.mass.ratio}(Color online) Total vibrational energy stored in the vibrating region, as a function of the mass of the electrode atom (black solid line: carbon; red dashed line: silicon; blue dot-dahsed line: germanium).}
\end{figure}

Except for modes between $153$ and $168$, the majority of the vibrational modes overlap with the phonon dispersions of the CNT electrodes.
This happens thanks to the same chemical character between hydrocarbons and carbon-based electrodes. 
If we were to consider organic molecules attached to a metal electrode such as gold or platinum, the large mass difference 
between atoms in the molecule and those in the electrode would make most of the molecular modes lie outside the electrode phonon dispersions. 
Therefore, most of the vibrational modes would have very small decay rates, significantly increasing the probability of the molecular junction to break down.

Together with Eqs. (\ref{RateEq2}), (\ref{eq:absorption_rate}) and (\ref{eq:emission_rate}), we can calculate nonequilibrium vibrational populations. 
Vibrational occupations will start to increase as the bias voltage exceeds their corresponding threshold voltages. 
While in the vicinity of the threshold voltage the vibrational populations increase linearly, nonlinear effects can appear at higher bias voltages. 
For low-energy modes vibrational populations monotonically increase with the bias; however, non-monotomic populations are observed for some of the high-energy modes. 
These trends are shown in Fig.\ref{cnt33-C6H4.noneq.phonons}, where vibrational populations for some of the highly excitable modes are illustrated. 
Mode $1$, which is low-energy, shows a monotonically increasing behavior. For the other high-energy modes, their populations increase, then decrease in a certain bias voltage range, and then start to increase again. 

One can hint at a qualitative explanation for this cooling behavior examining the local density of states in the device region, as recently reported in Ref.\cite{2010Romano}. 
As observed in Fig.\ref{cnt33-C6H4.dos}-(a), there is one resonant peak located at a energy higher than $\mu_{eq}$, which is the common Fermi level before the bias is applied. 
Figure \ref{cnt33-C6H4.dos}-(b) shows absorption and emission processes where electrons can exploit the resonant density of states for different bias voltages. 
Red and blue lines indicate absorption and emission processes respectively. 
For a very small bias none of the electron can access the resonant peak, as seen in Fig.\ref{cnt33-C6H4.dos}-(b)-(1). 
When the bias increases, the absorption process using the resonant peak starts to take place, but the electrons participating in the emission process cannot reach the resonance (See Fig.\ref{cnt33-C6H4.dos}-(b)-(2)). 
Therefore the absorption rate $A_{\lambda}$ becomes enhanced, so it may lead to a decrease in vibrational populations. 
For higher biases such that the resonant peak is located between the left and right chemical potentials, the emission process using resonance is activated, leading to enhanced emission rates, as shown in Fig.\ref{cnt33-C6H4.dos}-(b)-(3). 
When the bias increases more, another resonant emission process becomes possible, as shown in Fig.\ref{cnt33-C6H4.dos}-(b)-(4), while the resonant absorption process at which the electrons are reflected back to the left lead is prohibited due to Pauli blocking. 
As a result, the emission rate becomes more enhanced in comparison to the absorption one, and vibrational populations may increase again in this bias range. 
As an illustratative example, Fig.\ref{cnt33-C6H4.absorption.emission.rate} (b) shows the absorption and emission rates for mode $156$. 
One can clearly observe that the bias voltages for which the absorption and emission rates get enhanced are different: 
The slope of the absorption rate curve first increases around $0.4 \textrm{ V}$, but then decreases at $0.7\textrm{ V}$. 
By contrast, the emission rate linearly increases up to $0.8 \textrm{ V}$, and at a bias larger than $0.8 \textrm{ V}$ its slope also increases.
Difference between bias voltages at which absorption and emission rates become enhanced results in the observed cooling behavior in the intermediate bias range.

For low-energy modes whose energies are much smaller than the broadening of the resonant peak, the intermediate cooling regime may not appear distinctly. 
For example, see in Fig. \ref{cnt33-C6H4.absorption.emission.rate}-(a) the absorption and emission rates for mode $1$. 
Unlike mode $156$, the absorption and emission rates show quite similar dependence on the bias voltage, implying that a cooling behavior is not observed. 

Last, not every high-energy mode goes necessarily through a cooling down phase, since vibrational populations are determined by the interplay of absorption, emission, and decay rates. 
When the decay rate $\gamma_{\lambda}$ is larger than the difference between the absorption and emission rates $A_{\lambda} - E_{\lambda}$, 
the steady state solution can be approximated as
\begin{equation}
N_{\lambda} = \frac{n_{B}(\hbar \omega_{\lambda}) + E_{\lambda}}{\gamma_{\lambda} + A_{\lambda} - E_{\lambda}} \approx n_{B}(\hbar \omega_{\lambda}) + \frac{E_{\lambda}}{\gamma_{\lambda}}.
\end{equation}
In this case, the dependence of the vibrational population on the bias voltage becomes similar to that of the emission rate, and the cooling behavior does not appear. 

Finally, we would like to stress the importance of mass ratios between the conducting molecule and electrodes. 
As pointed out above, since the band width of bulk phonons in the electrodes gets narrower as the atomic mass of electrodes increases, 
a molecular junction connected to electrodes consisting of heavier atoms will have less opportunities to thermalize. 
To demonstrate this mass ratio effect, we calculate the total vibrational energy stored in the vibrating region 
by increasing the mass of the atoms in the carbon electrode to be e.g. silicon or germanium.
As shown in Fig.\ref{cnt33-C6H4.mass.ratio}, the junction with a larger mass ratio has more vibrational energy, and higher probability to break down due to heating effects. 

\section{Summary}
In this work we have described an \emph{ab initio} quantum transport approach based on maximally localized Wannier functions to calculate inelastic transport properties in the presence of electron-vibration interactions. 
In our implementation calculations are performed using the plane-wave DFT code \textsc{Quantum-ESPRESSO}, and \textsc{wannier90} for the transformation into MLWF and the construction of the Hamiltonian in the Wannier basis. 
Inelastic transport properties such as differential conductance and non-equilibrium vibrational occupations have been calculated within the Meir-Wingreen transport formalism based on NEGF. 
In particular, the electron-vibration self-energy is computed using either the SCBA or the LOPT. 
We have tested our implementation applying it to carbon-based molecular junctions: a benzene molecule connected to cumulene chains and a $(3,3)$ CNT-benzene-$(3,3)$ CNT junction. 

In the first system we have found the differential conductance change at bias voltages corresponding to the vibrational energies of active modes. 
Using multi-eigenchannel analysis based on scattering theory, we have identified that out-of-plane and in-plane vibrational motions lead to conductance jumps and drops respectively. 
This analysis on the simultaneous occurrence of conductance jumps and drops is consistent with recently reported calculation\cite{2009Kristensen}. 

In the second application, where cumulene wires are replaced by realistic CNT electrodes, we have focused on the decay rates of molecular vibrations and non-equilibrium vibrational populations. 
Small decay rates can be rationalized examining the band width of bulk phonons that molecular vibration can resonantly access, and the localization of molecular vibrations. 
Also, we have argued that a larger mass ratio between the conducting molecule and electrodes can enhance local heating and ultimately lead to the junction break down.
Thus organic molecular junctions connected to carbon-based electrodes or lighter metals may be more stable in comparison to heavy-metal electrodes (of course, strength of the anchoring chemical bonds will also play a role). 
We have also observed that nonequilibrium vibrational occupations for high-energy modes can be cooled down in an intermediate bias regime, which may be understood in terms of a resonant state in the device region. 
This observation agrees well with resonant cooling reported in other first-principle calculations\cite{2010Romano}.

These applications and detailed analysis, in a very good agreement with other first-principle studies performed with a local orbital basis set, 
confirm that our \emph{ab initio} approach to inelastic transport with plane-wave basis and Wannier functions could be applied to other realistic nanoscale junctions.

\begin{acknowledgments}
We are grateful to J. D. Joannopoulos, N. Bonini, D. Ceresoli, C.-H. Park, and X. Qian for many helpful discussions.
S. Kim acknowledges financial support from Samsung Scholarship. 
\end{acknowledgments}

\end{document}